# Data-driven Improved Sampling in PET

P. Galve, A. Lopez-Montes, J.M. Udías, S.C. Moore, J.L. Herraiz

[1] *Abstract–*Positron Emission Tomography (PET) scanners are usually designed with the goal to obtain the best compromise between sensitivity, resolution, field-of-view size, and cost. Therefore, it is difficult to improve the resolution of a PET scanner with hardware modifications, without affecting some of the other important parameters. Iterative image reconstruction methods such as the ordered subsets expectation maximization (OSEM) algorithm are able to obtain some resolution recovery by using a realistic system response matrix that includes all the relevant physical effects. Nevertheless, this resolution recovery is often limited by reduced sampling in the projection space, determined by the geometry of the detector.

The goal of this work is to improve the resolution beyond the detector size limit by increasing the sampling with data-driven interpolated data. A maximum-likelihood estimation of the counts in each virtual sub-line-of-response (subLOR) is obtained after a complete image reconstruction, conserving the statistics of the initial data set. The new estimation is used for the next complete reconstruction. The method typically requires two or three of these full reconstructions (superiterations). We have evaluated it with simulations and real acquisitions for the Argus and Super Argus preclinical PET scanners manufactured by SMI, considering different types of increased sampling. Quantitative measurements of recovery and resolution evolution against noise per iteration for the standard OSEM and successive superiterations show promising results. The procedure is able to reduce significantly the impact of depth-of-interaction in large crystals, and to improve the spatial resolution.

The proposed method is quite general and it can be applied to other scanners and configurations.

## I. Introduction

THE limited resolution of Positron Emission Tomography (PET) is one of its main drawbacks with respect to other imaging techniques such as Computerized Tomography (CT) or Magnetic Resonance Imaging (MRI). The resolution loss in PET is caused by a combination of different factors intrinsic to the technique [1]: positron range, non-collinearity, and radiation detection in the scanner. The first two are related to the radiation emission, and therefore they can be hardly improved. On the other hand, the geometry and configuration of modern state-of-the-art scanners are already optimized to get the best compromise between resolution, sensitivity, field-of-view (FOV), and cost. Therefore, it is difficult to improve the resolution of a PET scanner with any hardware modification, without affecting some of the other important parameters.

The size (length and width) of the scintillator crystals of the PET detectors is one of the most critical parameters of a PET scanner. Larger crystals increase the scanner sensitivity, but they worsen the resolution due to the depth-of-interaction (DOI) uncertainty [1]. The use of two or more layers of crystals (phoswich) reduces the DOI uncertainty [2], but requires more expensive detectors and electronics. The width of the crystals in a detector block (~4mm in clinical scanners and ~1.5mm in preclinical ones) also imposes an important limit to the resolution. This width is chosen based on a variety of factors such as the capability of the photomultiplier and electronics to differentiate the position of the events, or the size of the individual SiPMs, or the final cost of the detector blocks.

Iterative image reconstruction methods such as the ordered subsets expectation maximization (OSEM) algorithm may obtain some resolution recovery by using a realistic system response matrix that includes all the physical effects already mentioned [3, 7]. Nevertheless, this resolution recovery is often limited by the reduced sampling in the projection space.

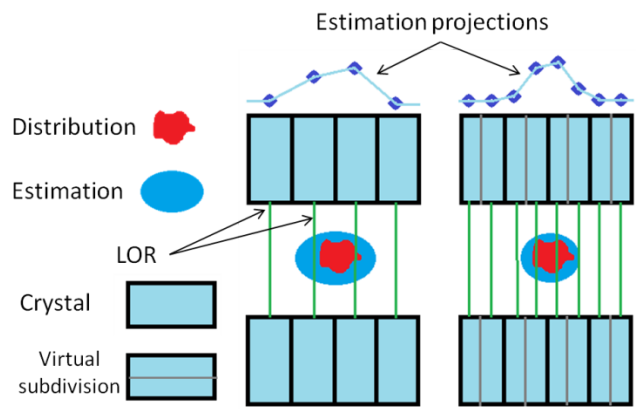

Fig. 1. Schematic diagram of the proposed data-driven interpolations.

In this work, we propose a method to further improve the resolution recovery in the image reconstruction process by iteratively refining the acquired data with improved sampling.

[1] Manuscript received November 15, 2017. This work was supported by Comunidad de Madrid (S2013/MIT-3024 TOPUS-CM), Spanish Ministry of Science and Innovation, Spanish Government (FPA2015-65035-P, RTC-2015-3772-1). This is a contribution for the Moncloa Campus of International Excellence. Grupo de Física Nuclear-UCM, Ref.: 910059. This work acknowledges support by EU's H2020 under MediNet a Networking Activity of ENSAR-2 (grant agreement 654002).

J. L. Herraiz is also funded by the EU Cofund Fellowship Marie Curie Actions, 7th Frame Program.

P. Galve is supported by a Universidad Complutense de Madrid, Moncloa Campus of International Excellence and Banco Santander predoctoral grant, CT27/16-CT28/16.

P. Galve, A.L. Montes, J.M. Udías and J.L. Herraiz were with the Grupo de Física Nuclear and Uparcos, Facultad de Ciencias Físicas, Universidad Complutense de Madrid, CEI Moncloa, Madrid 28040, Spain (pgalve@nuclear.fis.ucm.es).

S.C. Moore is with the Division of Nuclear Medicine, Brigham and Women's Hospital & Harvard Medical School, Boston, USA.

It is based on the fact that any standard image reconstruction process requires a series of linear interpolations due to the discrete sampling of both the image and the projection space. Improving the data sampling with subdivisions based on maximum-likelihood weights, yields data-driven interpolations, which are able to provide better resolution than standard methods (fig. 1).

The method was already applied for demultiplexing data in multiplexed SPECT [4]. In this case, we have used it to improve PET resolution by subdividing PET crystals in two different cases:
1. Reducing their length.
2. Reducing their width.

We have evaluated its performance with phantom data from two preclinical scanners. In both cases we obtain significant improvement of peak-to-valley ratio, recovery coefficients and resolution, whereas noise stays at similar values.

## II. Materials and Methods

### A. The superiterative method

The full algorithm of the proposed method comprises the following steps:

1. Standard OSEM reconstruction (fig. 2.1). We start with a PET acquisition with $N$ data elements $Y_i$. They could be elements of a sinogram, a line-of-response (LOR) histogram, or a list-mode data file. The image is represented by voxels $X_j$, and they are connected with the data via the system matrix with elements $A_{ij}$ as $Y_i = \sum_j A_{ij} X_j$. The first image $X_j^{k=0}$ is obtained by standard OSEM reconstruction.

2. Estimation of the augmented data sampling (fig. 2.2). Each crystal is virtually subdivided into two halves. For each initial LOR $Y_i$ there will be 4 different possible subLORs $Y_{i,h}^k$ ($h = 1...4$) connecting the centers of the subcrystals.

$$Y_{i,h}^k = Y_i \cdot W_{i,h}^k, \quad W_{i,h}^k = P_{i,h}^k / P_i^k, \quad P_i^k = \sum_h P_{i,h}^k \quad (1)$$

The weights $W_{i,h}^k$ correspond to a maximum-likelihood estimation as shown in [4,5], and they are obtained using the relative values of the projection $P_{i,h}^k$ in each subLOR with respect to the total projection in the LOR $P_i^k$, using the last reconstructed image $X_j^{k-1}$ (2).

$$P_{i,h}^k = \sum_j A_{(i,h)j} X_j^{k-1} \quad (2)$$

3. We reconstruct the image $X_j^k$ using the $4N$ data elements $Y_{i,h}^k$ and the standard OSEM algorithm (fig. 2.3).

We repeat steps 2-3 (we called each loop $k$ a superiteration) until convergence is reached. Typically, two or three superiterations are enough.

At step 2, the last image is used to generate the new data distribution between subLORs. This is the reason why the method is considered to be "data-driven": we use the whole acquired data to reconstruct the image, and the image is used to improve the precision of each LOR. Therefore, for a given LOR, the information in all the other LORs connected with it (i.e. sharing voxels with it) is used to improve the weighs of each subLOR.

It is important to point out that the proposed method simply redistributes the acquired counts, within the crystals and therefore it preserves the statistics of the acquired data.

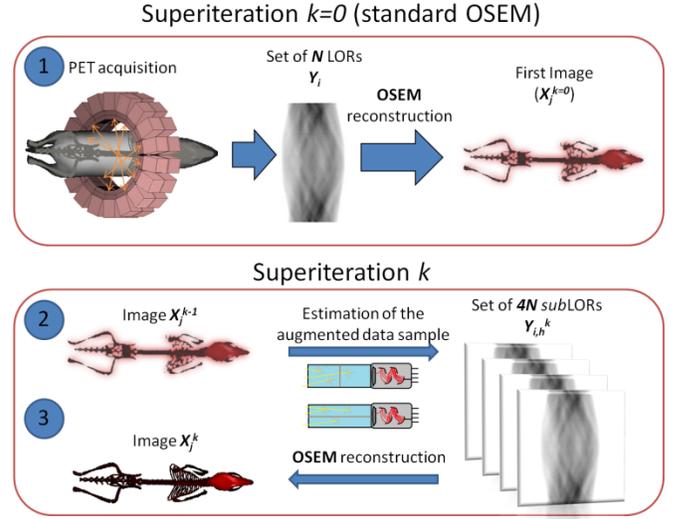

Fig. 2. Schematic diagram of the proposed method. We start with an initial acquisition, and reconstruct the image using the OSEM method (up, step 1). Once we have obtained the first image ($k=0$), each superiteration $k$ involves two steps: first, projection of the last image to calculate the augmented data sampling (step 2), and subsequent reconstruction of the $4N$ subLORs (step 3). Instead of reconstructing with the standard sampling, the object is reconstructed assuming smaller detectors with data redistributed based on the previous reconstructed image.

The method is quite general and the number of subdivisions, and the way the crystal is subdivided is not fixed. We have evaluated two possible cases:

- Data-driven DOI correction (fig. 3.a) – By sub-dividing the crystal in the longitudinal direction we can create a virtual Front and Back detector similar to one with phoswich [2].
- Data-driven Resolution Recovery (fig. 3.b) – By sub-dividing the crystal in the transverse direction we can increase the sampling and resolution of the image in the XY-plane.

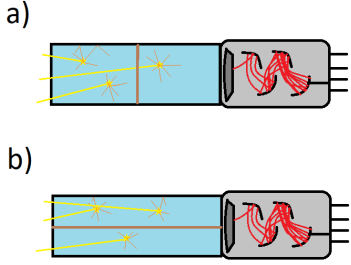

Fig. 3. Virtual subdivision for DOI correction (a) and transversal subdivision for resolution recovery in the XY-plane (b).

## B. Scanners and Data Acquisition

The scanners used in this work are the Argus PET/CT (Sedecal) and, a bigger similar one, the Super Argus PET/CT (Sedecal), both similar to the previous GE Healthcare eXplore Vista (General Electrics) [2]. These scanners are based on detector blocks of 13x13 coupled to double layer phoswich crystals LYSO-GSO.

For the Argus scanner, we first implemented the DOI correction (fig. 3.a), as this scanner has a smaller radius and the DOI effect is more notorious. We used a cold Derenzo Phantom filled with $^{18}$F-FDG. We compared the results obtained with the standard phoswich method used in this scanner [2] to correct the DOI, and the results obtained without phoswich and without phoswich but with the proposed method.

After that, we tested in the Argus scanner the transversal subdivision for resolution recovery shown in fig. 3.b. In this case, we used a $^{18}$F-FDG cardiac study with a rat.

For the Super Argus scanner we implemented the resolution recovery shown in fig. 3.b.

## I. RESULTS

We can see in fig. 4 a cold Derenzo Phantom acquired with the Argus scanner. When DOI information is disregarded, many artifacts come out, and resolution is considerably degraded (fig. 4.b), after the DOI correction is applied the image noise and resolution improves (fig. 4.c). In fact, at the corner of the green line profile shown below the images in fig. 4 we can see that the outer rod of 4 mm diameter is retrieved.

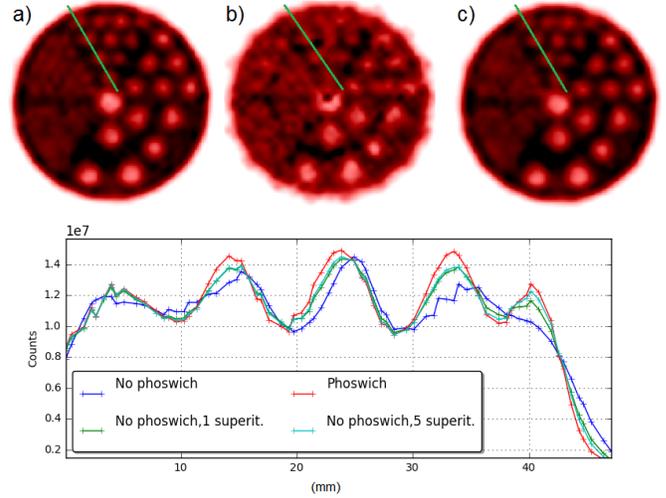

Fig. 4. Cold Derenzo Phantom, Acquired with the Argus scanner and reconstructed with 1 iteration and 50 Subsets of the OSEM algorithm. Images with and without phoswich (a and b), and without phoswich after 5 superiterations recovering DOI information (c). The line profile along the green line is shown at the bottom.

The result of the resolution recovery with the Argus scanner is shown in fig. 5, where we have applied the method to a rat acquisition. After the second superiteration, we achieved better resolution and spill-over reduction than the standard reconstruction, as it can be observed in the line profiles.

Results from a Super-Derenzo Phantom acquired with the SuperArgus scanner are shown in fig. 6. In this case, the effect of the method is difficult to be seen directly in the images. Nevertheless, a quantitative analysis along a line profile drawn across 1.5 mm diameter rods, shows an increase in the peak-to-valley ratio with respect to the standard reconstruction of 41±3% after the first superiteration and 54±3% after the second one.

Further measurements to test the method are shown in figs. 7-8 where the noise-resolution and noise-recovery evolution per iteration is plotted for OSEM and MAP-OSEM [6] algorithms over an NEMA NU4 Image Quality Phantom [8] (IQ phantom). It is clear that by using superiterations all the curves are improved.

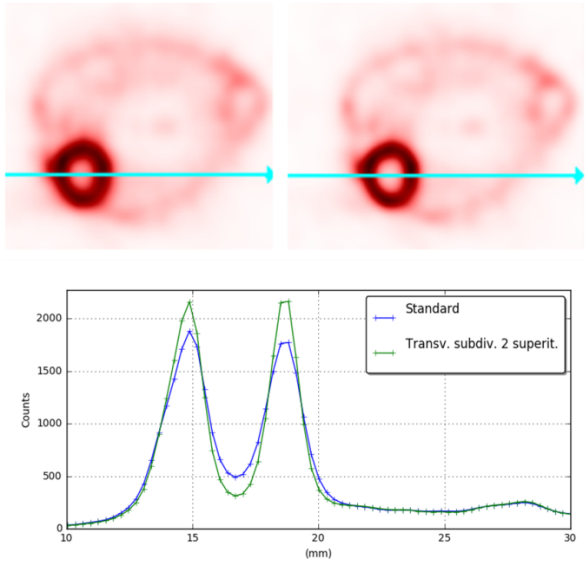

Fig. 5. Transverse view of a rat heart injected with FDG and acquired with the Argus scanner. (A) Standard OSEM reconstruction (10 iterations, 10 subsets). (B) After 2 superiterations with same parameters, recovering transversal information. The line profile along the blue line, crossing the heart, is shown below the images. A significant improvement in the resolution and reduction of the spill-over of the myocardium activity into the left ventricle is seen.

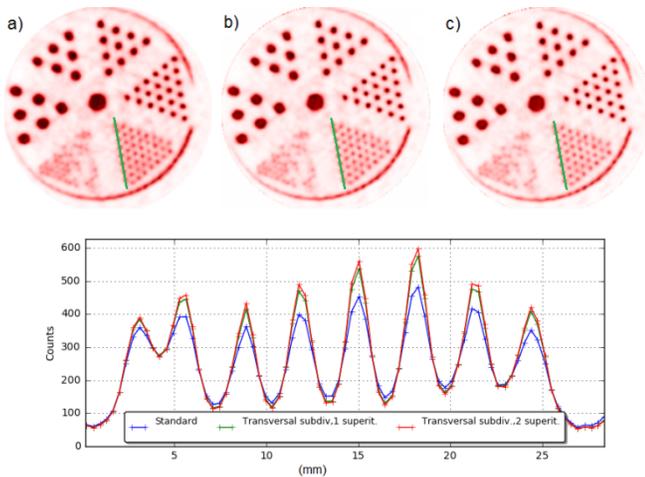

Fig. 6. Reconstructed Super-Derenzo acquired with the SuperArgus scanner. (a) Standard reconstruction OSEM (20 iterations, 5 subsets) and after 1 and 2 superiterations recovering transversal information (b and c). The line profile along the green line is shown at the bottom.

### B. Noise-Resolution curves

We measured resolution by differentiation of the line profile at the edge of a rod of 4 mm diameter, and subsequently fitting a Gaussian curve for which we show the FWHM value, and noise was measured over a uniform region of 4 cm diameter and 1.5 cm height. In the upper graph, we observe the standard OSEM curve saturates after ~200 image updates (20 iterations of 10 subsets), superiterations achieve a resolution improvement (lower FWHM values) without noise increment. To prove that the method also behaves like standard OSEM for a higher number of iterations, we ran 40 iterations for the third superiteration, showing saturation for submilimetric resolution.

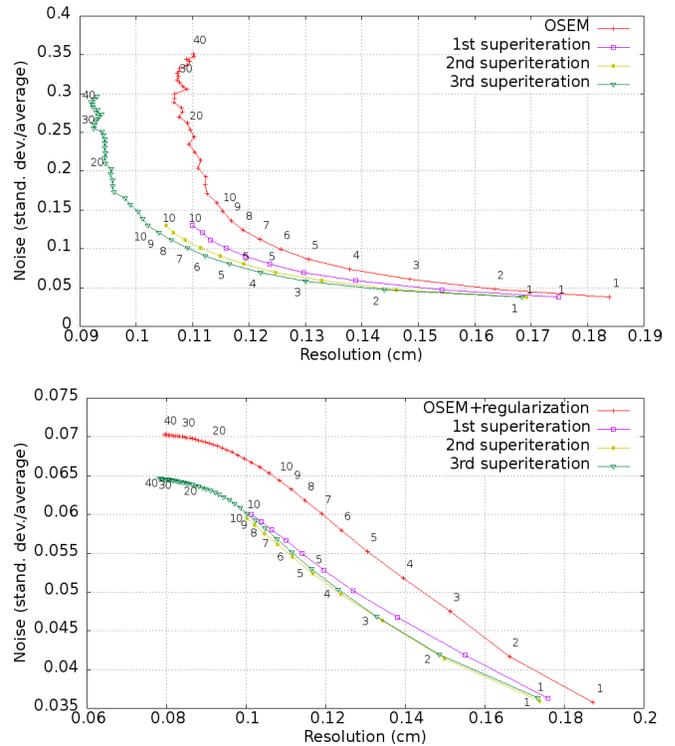

Fig. 7. Noise-resolution curves for an IQ phantom acquired with the preclinical Argus scanner reconstructed using standard OSEM (10 subsets) (up) and MAP-OSEM with 10 subsets, $\beta = 0.08$ (down). We applied the transverse information recovery method, projecting always the 10th iteration image to compute the next superiteration weights. The number of iterations of each point is indicated.

We observe similar results when MAP-OSEM is applied (fig. 7, down), although the regularization parameter allows lower noise levels and slower convergence. As a result, we can achieve better resolution since it is not limited by the uncontrolled noise increase found for OSEM.

### C. Noise-Recovery curves

For the recovery measurements, we considered the average activity accounted inside a 4 mm diameter and 1.2 cm height region of interest (ROI), and the average activity inside a uniform volume. (Noise was equally measured as for resolution; note we used exactly the same measurements)

The noise-recovery curves in fig. 8 show a similar behavior to noise-resolution against iterations and superiterations. Recovery coefficients increase with noise and further image updates, up to a certain limit. In contrast, the MAP-OSEM algorithm does not achieve better recovery for this magnitude while superiterations keep increasing recovery.

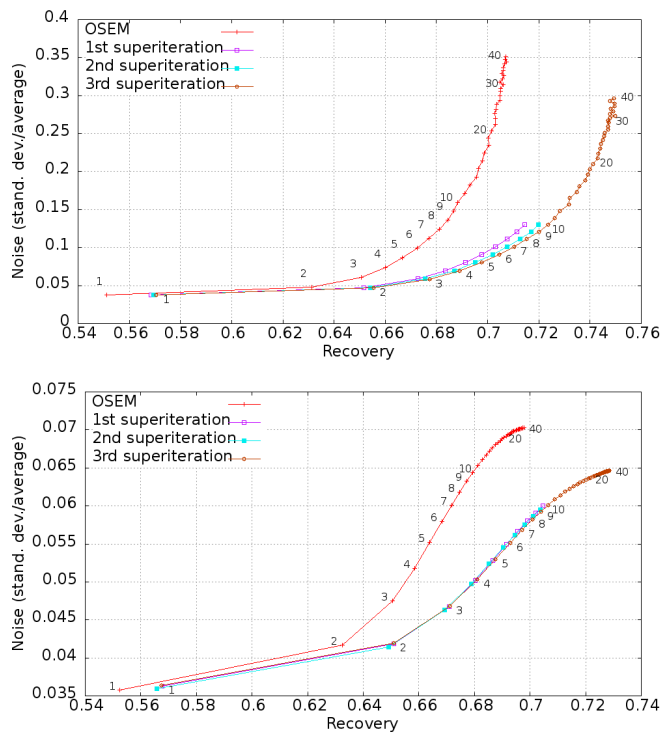

Fig. 8. Noise-recovery curves for an IQ phantom acquired with the preclinical Argus scanner reconstructed using standard OSEM (10 subsets) (up) and MAP-OSEM with 10 subsets, $\beta = 0.08$ (down). We applied the transverse information recovery method, projecting always the 10th iteration image to compute the next superiteration weights. The number of iterations of each point is indicated.

## II. Discussion and Conclusions

The method is quite general, as it has been successfully applied to different kind of subdivisions and scanners. Several proofs of its potential effectiveness to reduce artifacts and increase image quality have been shown (figs. 4-5). A peak-to-valley ratio increase of 41±3% in 1.5mm rods with respect to the normal OSEM method has been achieved. Besides we also measured noticeable quantitative resolution and recovery improvement without noise increase (figs. 7-8).

Comparison of plain OSEM results with three superiterations just before the noise saturation knee, we see an improvement of 15% of the resolution and an increase of the RC by 5%. These results are also found when OSEM-MAP reconstructions are compared to superiteration+MAP ones. The resolution increase has to be attributed to the method. Superiteration introduces additional degrees of freedom in the SRM that are consistently relaxed until a converged image, with projections more consistent with data, is obtained.

It is important to remark that the superiterative method increases several times the reconstruction time, but this is a minor concern with current high-performance computers and GPUs. Computation time per iteration estimated was ~2 times higher when we are working with the *4N* data set (recall we have 4 times the number of LORs when a superiterative reconstruction is in progress).

Further studies on the impact of the method on all the relevant parameters of image quality are ongoing to prove the improvement of resolution against noise of the method.